\documentclass[12pt,preprint]{aastex}

\shorttitle{Enhanced X-ray and Infrared emission from AXP 1E 2259+58}
\shortauthors{Ertan, G\"{o}\u{g}\"{u}\c{s} $\&$ Alpar 2003}

\begin{document}


\title{X-Ray and Infrared Enhancement of \\
 Anomalous X-ray Pulsar 1E 2259+58 }


\author{
{\"{U}. Ertan\altaffilmark{},  
E. G\"{o}\u{g}\"{u}\c{s}\altaffilmark{}
$\&$ M. A. Alpar\altaffilmark{}}}
\affil{Sabanc{\i} University, FENS, Orhanl{\i}$-$Tuzla 34956 
{\.I}stanbul/ Turkey}
 

\begin{abstract} 
The long term ($\sim 1.5$ years) X-ray enhancement and 
the accompanying infrared enhancement 
light curves of the anomalous X-ray pulsar 1E 2259+58 following the 
major bursting epoch can be accounted for by the 
relaxation of a fall back disk that has been pushed back by a 
gamma-ray flare. 
The required burst energy estimated from the results
of our model fits is low  
enough for such a burst to have remained below the detection limits. 
We find that an irradiated disk model with a low irradiation efficiency 
is in good agreement with both X-ray and infrared data.   
Non-irradiated disk models also give a good fit to the X-ray light 
curve, but are not consistent with the infrared data for the first 
week of the enhancement.

\end{abstract}

\def\la{\raise.5ex\hbox{$<$}\kern-.8em\lower 1mm\hbox{$\sim$}}
\def\ga{\raise.5ex\hbox{$>$}\kern-.8em\lower 1mm\hbox{$\sim$}}
\def\be{\begin{equation}}
\def\ee{\end{equation}}  
\def\ba{\begin{eqnarray}}
\def\ea{\end{eqnarray}}  
\def\be{\begin{equation}}
\def\ee{\end{equation}}  
\def\ba{\begin{eqnarray}}
\def\ea{\end{eqnarray}}  
\def\m{\mbox}
\def\d{\partial}
\def\R{\right}
\def\L{\left}
\def\a{\alpha}
\def\Mdot*{\dot{M}_*}
\def\Mdotin{\dot{M}_{\mbox{in}}}
\def\Mdot{\dot{M}}
\def\Lin{L_{\mbox{in}}}
\def\Rin{R_{\mbox{in}}}
\def\Rout{R_{\mbox{out}}}
\def\Ldisk{L_{\mbox{disk}}}
\def\dEb{\delta E_{\mbox{burst}}}
\def\dEx{\delta E_{\mbox{x}}}
\def\Bb{\beta_{\mbox{b}}}
\def\Be{\beta_{\mbox{e}}}
\def\dMin{\delta M_{\mbox{in}}}
\def\dM*{\delta M_*}
\def\Teff{T_{\mbox{eff}}}
\def\Tirr{T_{\mbox{irr}}}





\section{Introduction}

Anomalous X-ray pulsars (AXPs) and
soft gamma ray repeaters (SGRs) are neutron stars 
which are characterized by 
their persistent X-ray luminosities 
($L_{\m{}}\sim 10^{34} - 10^{36}$ erg s$^{-1}$)
well above their rotation powers (see Mereghetti 2002 and 
Hurley 2000 for reviews of AXPs and SGRs respectively). Another remarkable property of these
sources is the clustering of their spin periods to a very narrow range 
($P \sim $ 5 - 12 s). 
There is no observational evidence 
for binary nature in these systems. They exhibit
short ($\la $ 1 s),  super-Eddington      
 ($\la 10^{42}$ erg s$^{-1}$) bursts. 
Burst repetition time scales vary 
from seconds to years.
In about three decades of observations of these sources, three giant flares 
were observed from the 
SGRs 0526-66 (Mazets et al. 1979), 1900+14 (Hurley et al. 1999) 
and 1806-20 (Palmer et al. 2005). 
These giant flares are characterized by an initial hard spike with a 
peak luminosity  $\ge 10^{44}$ erg s$^{-1}$
 which lasts a fraction of a
second. The hard spike is followed by an oscillating tail that
decays in a few minutes. Assuming isotropic emission, the fluence of
the entire giant flare is $\ga 10^{44}$ ergs 
(Hurley et al. 1999; Feroci et al. 2001; Mazets et al. 1999). 
Following the August 27$^{th}$ 1998 giant burst of SGR 1900+14, 
the persistent X-ray flux increased by a factor of $\sim 700$ 
to a peak luminosity $\sim 10^{38}$ erg s$^{-1}$.  
The subsequent  decay is a power law with index $\sim 0.7$ (Woods et
al. 2001).
The three other, less luminous X-ray enhancements of SGR 1900+14 
were also preceded by gamma-ray flares.    
An X-ray enhancement was also observed from AXP 1E 2259+58. Although no strong
burst preceding the enhancement was reported, it is possible that such a burst
could have been missed (Woods et al.2004). Out of six known AXPs including 
1E 2259+58,  five sources were 
detected in the IR band (Hulleman et al. 2001; Israel et al. 2002; Wang $\&$
Chakrabarty 2002; Kaspi et al. 2003; Israel et al. 2003; Hulleman et al. 2004;
Israel et al. 2004). One of them, 4U 0142+61, was also observed in the 
optical (R) band (Hulleman et al. 2000, Dhillon et al. 2005). The optical flux
from AXP 4U 0142+61 has a pulsed component with a pulsed fraction 
$(\sim 0.27)$ much higher than the observed pulsed fraction $(\sim 0.05)$
in X-rays (Kern $\&$ Martin 2002). {\bf Observations of transient AXP XTE J1810-197 separated by a few months showed a correlated decrease (by a factor $\sim 2$) in  X-ray and IR luminosities (Rea, N et al. 2004, Israel et al 2004). IR radiation from this source was proposed to be radiated from a passive disk irradiated by the X-rays from the neutron star Rea et al. (2004). In this model, the source of the X-rays is assumed to have magnetic origin, as proposed by the magnetar models, rather than accretion onto the neutron star.} The X-ray enhancement of AXP 1E 2259+58 was also accompanied by an enhancement in the IR luminosity (Kaspi et al. 2003). Long
term ($\sim 1.5 $ years) IR and X-ray flux evolutions of this source exhibited a
correlated decay behavior (Tam et al. 2004).      

The magnetar model (Duncan $\&$ Thompson 1992, Thompson $\&$ Duncan 1995)
can explain the energies and the super-Eddington luminosities of the
normal and the giant bursts of these sources. In these models, the source
of the bursts 
is the huge   magnetic energy release from inside the neutron stars. The 
persistent X-ray emission was suggested to originate from the  
magnetic field decay (Thompson $\&$ Duncan 1996). The spin down 
torque is provided by the magnetic dipole radiation. Magnetar models have no
detailed predictions for the observed optical and infrared (IR) emission 
from AXPs and cannot explain
the period clustering of AXPs and SGRs without assuming special conditions 
(Colpi, Geppert $\&$ Page 2000). 

Fall back disk models (Chatterjee, Hernquist, 
$\&$ Narayan 2000; Alpar 2001; Marsden et al. 2001) explain the 
period clustering of these sources using conventional dipole magnetic fields of
young neutron stars with $B_* \sim 10^{12} - 10^{13}$ G on the surface of the neutron
star within
the time scales of the ages of their associated supernova remnants 
(Chatterjee $\&$ Hernquist 2000; Ek\c{s}i $\&$ Alpar 2003). 
In fall back disk models, the source of the X-ray luminosity is the accretion 
onto the neutron star from the accretion disk. In earlier work (Ertan $\&$ Alpar 2003), 
we showed, by means of numerical fits to data, that 
the X-ray enhancement of the SGR 1900+14
following its giant flare can be reproduced by the relaxation of the inner disk
subsequent to an initial push back caused by the flare. 
We estimated that part of the pushed back inner
disk piles up and remains bound creating the post burst initial conditions of the
disk, while the remaining part escapes from the system. 
The estimated energetics and the amount of escaping ejecta 
can account for the observed radio enhancement accompanying
the X-ray enhancement of the SGR 1900+14 (Ertan $\&$ Cheng 2004b). An important
limitation for the models is the observed optical and IR emission from the AXPs. 
It was shown that the observed pulsed component of AXP 4U 0142+61 
can be explained by both the magnetar outer gap models and the disk-star 
dynamo models (Ertan $\&$ Cheng 2004a). Furthermore, unpulsed optical/IR emission
from the AXP 4U 0142+61 is in agreement with the expectations of a standard thin disk
model (Ertan $\&$ Cheng 2004a). In the disk models, the inner radius of the disk
is consistent with a dipole component of the magnetic field, with strength 
$\sim 10^{12}$ G on the stellar surface. The origin of energetic bursts require
magnetar strength surface fields. It could be the case that the burst involves
surface magnetar fields in the higher multipole components, while a disk
standing at inner radius set by a $10^{12}$ G surface dipole field defines the
rotation period of the star in a near-equilibrium state.      
It seems that such an hybrid model can explain almost all the observational 
behaviors of the AXPs and SGRs. 

{\bf The enhancement energetics of the SGR 1900+14 and the AXP 1E 2259+58 are
different. While the SGR 1900+14 was observed only in X-rays, we now have
contemporaneous X-ray and IR enhancement data for the AXP 1E 2259+58. In this
work, our aim is to answer: (i) Is it possible to explain the X-ray 
enhancement of the AXP 1E 2259+58 by using a similar pushed back disk model 
applied to the SGR 1900+14? (ii) Is this model capable of reproducing both the 
X-ray and the IR data of the AXP 1E 2259+58?  In answering these
questions, we test both irradiated and non-irradiated disk models. We estimate 
the required burst energy to build up post-burst initial inner disk
conditions for the enhanced radiation and check whether such a burst could have remained  below detection limits.}

The 2-10 keV X-ray flux data of AXP 1E 2259+586 were taken from Woods    
et al. (2004). They have monitored the source with the Rossi X-ray    
Timing Explorer / Proportional Counter Array, and determined the
pulsed X-ray intensity of the source over a large time baseline. Using       
pointed XMM-Newton observations, they derived a conversion factor between    
the pulsed intensity and unabsorbed flux for this source. A detailed    
description of this conversion is given in Woods et al. (2004).       
    
The infrared measurements of 1E 2259+586 were obtained from Kaspi
et al. (2003),  Israel et al. (2003) and Tam et al. (2004). 
The first two observations
were made with the Near-Infrared Imager (NIRI) at the Gemini North telescope at
about 3 and 10 days after the onset of the bursting activity. In these
pointings, the $K_s$ magnitudes of the source were found to be 20.36$\pm$0.15
and 21.14$\pm$0.21 respectively (Kaspi et al. 2003). The third infrared
measurement, carried out with the Canada-France-Hawaii Telescope, gave 
the $K^{\prime}$ magnitude to be 21.31 $\pm$ 0.24 (Israel et al. 2003).
The last two observations were also made with NIRI and resulted with $K_s$ 
magnitudes of 21.66$\pm$0.11 and 21.54$\pm$0.05 respectively. 
We converted these magnitudes into physical flux units using the method 
described in Tam et al. (2004). 
Details of the irradiated and non-irradiated disk models are given in \S 2. 
The results are discussed in \S 3. We summarize our conclusions in \S 4.

\section{The Numerical Model}

All four observed 
X-ray enhancements of the SGR 1900+14 were preceded by strong gamma-ray flares.
No gamma-ray burst was reported 
before the onset of the X-ray enhancement of the AXP 1E 2259+58. It is 
possible that such a burst could have remained below the detection limits. 
We assume that the X-ray enhancement of the AXP 1E 2259+58 was preceded 
 by a missed gamma-ray burst. We discuss the burst fluence required to build up the 
 post burst initial disk conditions and the observational limits in \S 3.      
 In earlier work
(Ertan $\&$ Alpar 2003), we showed that the X-ray enhancement evolution of the
SGR 1900+14 following the August 27 giant flare can be accounted for by the
relaxation of the inner disk matter that has been pushed back by the flare. 
We apply the same model to AXP 1E 2259+58 
assuming the physical mechanism giving rise to the X-ray enhancement is the same. 
Our aim is to explain the observed X-ray enhancement together with 
the accompanying IR enhancement data. In our model, the source of the 
X-rays is accretion onto the NS, while the IR emission originates from the
accretion disk. We investigate the IR emission for  
both irradiated and non-irradiated disk models. 
 
In our numerical model, we take the model  
functional form of the post burst initial mass
distribution to be the same as applied to SGR 1900+14 (Ertan $\&$ Alpar 2003). 
The part of the inner disk matter pushed back by the flare energy 
which remains bound to the system is represented by a Gaussian 
$\Sigma(R,t=0) = \Sigma_{\m{max}}~ exp \left[-\left(\frac{R-R_0}{\Delta R}
\right)^2 \right]$ centered at the radius $R_0$.  
 Added to this
Gaussian is a surface density profile in the form 
$\Sigma = \Sigma_0 (R_0/R)^p$ which represents 
 the extended outer disk. The inner disk radius $\Rin$ where the inflowing disk
 matter is stopped by the magnetic pressure is kept constant.  
 We numerically solve the disk diffusion equation for the surface density 
as described in detail in Ertan $\&$ Alpar (2003).
We use the $\a$ prescription of the kinematic viscosity 
$\nu = \a c_{\m{s}} h$ (Shakura $\&$ Sunyaev 1973) where 
$c_{\m{s}} = k T_{\m{c}} / \mu m_{\m{p}}$ is the
local sound speed, $T_{\m{c}}$ the local disk midplane temperature, 
$h = c_{\m{s}}/\Omega_{\m{K}}$ the pressure scale height of the disk and  
$\Omega_{\m{K}}$ the local Keplerian angular velocity of the disk. 
The viscosity parameter $\a= 0.08$ and the outer disk radius 
$R_{\m{out}}= 10^{12}$ cm are kept constant throughout the calculations. 
The chosen $R_{\m{out}}$  is large enough that it does not 
effect the results.

\subsection{X-ray Emission from the Neutron Star Surface }  

For a thin disk, the total disk luminosity is 
$\Ldisk= G M \Mdotin / 2 R_{\m{in}}$, and most of this emission
comes from the
inner disk. Here, $\Mdotin$  is the mass inflow rate arriving at the inner disk 
radius $\Rin$ and M is the mass of the neutron star, 
which we take to be $1.4  M_{\odot}$.
The accretion luminosity from the NS surface,  
$L_*= G M \Mdot_*/R_*$, determines the observed luminosity in the X-ray
band.  Assuming  most of 
the X-ray flux from the source 
is emitted in the observation band (2-10 keV),
we take the observed X-ray luminosity to
represent the total luminosity $L_*$ of the neutron star.  
The evolution of the X-ray luminosity depends on both $\Mdotin(t)$   
and its fraction  accreted onto the neutron star $f= \Mdot_*/\Mdotin$  
where $\Mdot_*$ is the mass accretion 
rate onto the star. The X-ray luminosity can be written as 
$L_{*} = 2 f (R_{\m{in}}/R_*) L_{\m{disk}}$.  
We compare the energy flux   
$F_{*} \sim L_{*} /(4 \pi d^2)$ with the X-ray data 
for the model fits. We take the distance of the source 
$d=5$ kpc (Hulleman et al. 2000).

\subsection{Infrared Emission from the Non-irradiated Disk}

For a thin, non-irradiated accretion 
disk, the source of the disk blackbody emission   
is the  viscous dissipation inside the disk. The dissipation rate $D$ per unit area sets the effective temperature and flux from both surfaces of the disk   
\be
D =\frac{9}{4} \nu \Sigma \Omega_{\m{K}}^2= 2 \sigma T_{\m{eff}}^4. 
\label{4}  
\ee  
The surface density and the corresponding dissipation rate profile at a 
given time is obtained by numerically 
solving the diffusion equation  (see Ertan $\&$ Alpar 2003 for details). 
The effective temperatures are 
calculated for each radial grid point.     
We obtain the model IR luminosity by integrating the blackbody fluxes 
along the spatial grid points emitting in the observational IR band 
{\bf ($K_s\, \lambda=2.15 \mu$m, width=0.31$ \mu$m). The blackbody temperature which
mainly emits in this band is $k\Teff\sim 0.2$ eV.}          
For the model fits,  
we relate the model IR luminosities to the observed flux  
by  $F_{IR} \sim L_{IR} \cos i/(4 \pi d^2)$ 
where $i$ is the inclination angle
between the normal of the disk plane and the line of sight of the observer.  

The model puts an upper limit to the 
fraction $f$ of the disk mass flow rate 
 which is accreted onto the neutron star. 
This upper limit is obtained by setting $\cos i =1$. 
For the  model given in Fig. 1, $f= 0.03$  and $\cos i= 0.9$. 
The other model parameters are given in
Table 1. It is seen in Fig. 1 
that the non-irradiated disk model is consistent with the long term evolution of the 
IR data ($\sim 1.5$ years) but not for the first two weeks.

\subsection{Infrared Emission from the Irradiated Disk}

A likely reason for this discrepancy is the irradiation of the disk by the X-rays from the neutron star. 
For a disk irradiated by the X-rays from the neutron star, the irradiation
flux through the disk surface must be added to the dissipative energy flux 
in Eq. 1. The X-ray irradiation flux can be written as 
\be
F_{irr}=\sigma T_{\m{irr}}^4 = C  \frac{\Mdotin c^2}{4 \pi R^2}
\label{6}   
\ee
where
\be
C= \eta (1-\epsilon) \frac{H_{\m{irr}}}{R} \left(\frac{d\ln H_{\m{irr}}}
{d \ln R}-1\right)
\label{7}
\ee 
(Shakura $\&$ Sunyaev 1973), $\eta$ is the efficiency of the 
conversion of the rest mass energy into X-rays, $\epsilon$ the X-ray albedo 
of the disk face, $R$ the radial coordinate and $H_{\m{irr}}$ the pressure scale height of the disk which 
should be calculated including the effect of the X-ray irradiation 
(Dubus et al. 1999). The disk thickness to 
radial distance ratio $H/R$ is roughly constant along the disk. 
The parameter $C$ can vary in a large interval especially due to the uncertainty 
on the disk albedo $\eta$. For a point irradiation source, estimates are usually 
in the range $10^{-4} - 10^{-3}$ (Tuchman et al. 1990, de Jong at al. 1996,
Dubus et al. 1999).   
Using  $C$ as a free parameter, we calculate the effective temperature for 
the irradiated disk as 
\be   
\sigma T_{\m{eff}}^4= \frac{D}{2} + F_{\m{irr}}.
\label{8}
\ee
{\bf Eq. 2 shows that $\Tirr\propto R^{-1/2}$, while $\Teff\propto R^{-3/4}$ for
a non-irradiated disk. For small radii viscous heating dominates over the
X-ray 
heating. Beyond some critical radius $R_c$, depending on the
irradiation strength, the main source of the emission from the disk surface is
the reprocessed X-rays for an irradiated disk. The critical radius $R_c$
 can be estimated by equating the irradiation flux $F_{irr}$ 
 to the dissipative energy flux $D/2$ (Eq. 1). Assuming that the mass 
 inflow rate at a given time is constant along the disk, we obtain} 
\be
R_c=\frac{3}{2} \frac{GM_*}{C c^2}\simeq f \left(\frac{10^{-4}}{C}\right) 
3\times 10^9 \m{cm}
\ee  

We follow the same method as that of the non-irradiated disk 
model to calculate the total IR luminosity in the observational 
IR ($K_{\m{s}}$) band. 
In the vertical disk analyses, the X-rays are usually assumed 
to be absorbed in a thin layer of the disk surface. The irradiation modifies 
the vertical temperature profile and 
the effective temperature of the disk without 
significantly affecting the disk mid-plane temperatures that determine the
kinematic viscosity (Dubus et al. 2001). Therefore, X-ray irradiation 
does not change the evolution of the disk mass transfer rates 
as long as the disk remains at the same viscosity state (see \S 3). 
The irradiated disk model cannot constrain the $f$ parameter since there 
is the extra free parameter $C$ .   
The model X-ray and IR flux curves given in Figures 2 $\&$ 3, were 
obtained with $C=2\times 10^{-4}$ and $f=0.26$ for  $\cos i=0.76$. 
When we take $f$ near unity with $\cos i=0.76$, a similar model fit is obtained
with $C\simeq 5\times 10^{-5}$. 
{\bf The annular disk section radiating in the $K_s$ band is at a radial
distance $R\sim 2\times 10^{11}$ cm at the beginning of the enhancement, and
shrinks slowly to  $R\sim 5\times 10^{10}$ cm along the decay phase, remaining in the irradiation dominated part of the disk.} 
Figures 2 $\&$ 3 show that this particular irradiated disk model is in agreement with 
both the X-ray and the IR data.       

\section{Results and Discussion} 

The model X-ray and IR light curves are presented for the
non-irradiated (Fig. 1) and irradiated (Figures 2 $\&$ 3) disk models. The corresponding model parameters
are given in Table 1. The differences between the initial surface density 
distributions of the best fitting models are their amplitudes and a small change 
in the power-law index $p$ of the extended disk profiles. 
In the irradiated disk model, since the effective temperatures are mainly 
determined by the irradiation flux rather than the viscous flux,    
the amplitude of the initial surface density distribution is less than that of the 
non-irradiated disk model (by a factor $\sim 5$). The time evolution of the mass
accretion rate $\Mdot$ onto the neutron star which is observed as the X-ray
light curve is similar for all different models of the disk. 
For a given initial mass 
distribution, the evolution of the mass flow rate 
$\Mdot$ through the disk is the same for both models, 
as the irradiation has no significant effect on the disk midplane temperatures. 
In the hydrogen disk models employed  to account for the observed 
outbursts of dwarf novae and soft X-ray transients, the disk has two states. 
The viscosity parameter $\a$ 
is large (small) when 
the effective temperature of the disk is high (low) compared to a critical
temperature reflecting the ionization of hydrogen ($\sim 10^4$ K). 
A local disk instability 
can lead to an outburst (quiescence) if the resultant heating (cooling) wave 
can take all  or most of the disk to the high (low) viscosity state.   
The inclusion of the X-ray irradiation in models modifies significantly the 
critical accretion rates at which these instabilities occur. Nevertheless, 
for a given stable state, the effect of irradiation on disk mass flow 
evolution is small. 
{\bf The critical mass flow rates and the critical temperatures for 
disk instabilities depend on the
exact composition of the disk material which determines the opacities 
in the disk (Menou et al. 2002). They use carbon - oxygen composition to discuss
disk instabilities. This composition is uncertain for 
AXP and SGR disks. In all our calculations, we use the electron scattering opacity 
as we previously applied for SGR 1900+14.}   
During our simulations we kept the viscosity parameter constant at the value 
$\a=0.08$ which 
is characteristic of the hot states in hydrogen disk models.    
We also tested disk models with both high and low viscosity states occurring
(bimodal) going through the hydrogen ionization temperature. 
We used $\a$ parameters similar to those of hydrogen disk models 
($\a_{hot}\sim 0.1$  and  $\a_{cold}\sim 0.01 - 0.05$). 
{\bf For a critical effective
temperature $\sim10^4$ K,} both X-ray and  
IR model curves deviate from the observed trend of the 
light curves significantly 
following the instability. The X-ray enhancement light curve of the SGR 1900+14 
can also be reproduced by using a constant $\a$ parameter $\a \sim 0.1$ 
(Ertan $\&$ Alpar 2003). {\bf Secular decay of the light curves imply that 
in the accretion regime indicated by the observed X-ray luminosities of the
AXPs and SGRs, fall-back disks do not experience a global disk instability.}

The rise and early decay phase (the first few days of enhancement) of 
the X-ray luminosity are determined 
by the release of the Gaussian surface density distribution representing the 
post burst pile up. In this phase, X-ray model curve evolution depends mainly on the 
amount of mass included in the Gaussian and the distance between the 
Gaussian center ($R_0$)
and the disk's inner radius, rather than the detailed Gaussian parameters. 
{\bf The two model light curves seen in Fig. 2 are for two
different initial Gaussian surface density distributions with the same 
extended (outer) disk profile. The constant inner disk radius $\Rin$ was 
taken near the Alfv$\acute{e}$n radius for a dipole magnetic field with strength 
$\sim 10^{12}$ Gauss on the stellar surface and 
$\Mdotin\sim 10^{15}$ g s$^{-1}$.
Similar fits can also be obtained for larger $\Rin$ and $R_0$. 
The model fits do not constrain the dipole magnetic field strength.}
For the first few weeks, the decay phase does not strongly depend on the chosen 
power-law index $p$ of the extended model disk surface density profile, 
provided it remains roughly around its best fit value for the overall 
evolution curve. About a few weeks after the onset 
of the enhancement, light curve evolution becomes sensitive to the power
index $p$. For a steady disk, $p$ is expected to be around 3/4, taking the 
run of the midplane temperatures as $R^{-3/4}$. 
This value of $p$ is close to our best fit  $p$ parameters 
for both irradiated and non-irradiated disk models (0.65 and 0.70). 
 {\bf In Fig. 3, we illustrate three irradiated disk  
model light curves for different $p$ values (0.60,0.65 and 0.70), 
keeping the Gaussian parameters constant. 
Irradiated and non-irradiated models are successful in explaining the 
long term X-ray enhancement data.
The non-irradiated disk model can reproduce the IR 
data except for the first two weeks with a low upper limit $f\le 0.03$ 
for the accretion rate to disk mass flow rate ratio $f$.   
The irradiated disk model gives a better fit to the overall IR data 
without constraining $f$, due to the uncertainty in the efficiency of the 
X-ray irradiation for the AXP and SGR
disks which is likely to be different from the LMXB disks. 
For LMXBs optical to X-ray luminosity ratios are much higher than expected optical
luminosity from the intrinsic dissipation of the disk and the X-ray luminosity  
from the neutron star surface. While this indicates that the optical flux of LMXBs
are from an X-ray irradiated disk, theoretical calculations of the disk thickness
profile show that irradiation must be indirect because of the self-screening of the
disk (See e.g. Dubus et al. 1999). Most likely source of the indirect emission is a
hot scattering corona above and below the central disk region. The hot corona can be
fed by the thermally unstable, optically thin surface layers of the hot innermost
disk. Is it plausible to expect that the SGR and AXP disks have the same 
X-ray irradiation efficiency
as that of LMXBs? This is not likely because the stronger magnetic 
fields of SGRs and AXPs cut the disk at
radii two to three orders of magnitude larger than the inner disk radii of LMXBs.
For the irradiated disk model, we initially fix the parameter $f= \Mdot*/\Mdotin$
and try to obtain a good fit to the X-ray data points first. Then, we look for the
best fit to the IR light curve tracing the irradiation strength $C$. The
irradiated disk model curves presented in Figures 2 and 3 are obtained with 
$f\simeq 1/4$, $\cos i\simeq 3/4$ and $C= 2\times 10^{-4}$. We see that
fits with similar quality can be obtained for 
$f C \cos i \sim 4\times 10^{-5}$.}           
       
The gamma-ray burst energy required to build up the initial conditions of 
our models can be estimated by assuming the post burst pile up represented by 
Gaussian surface density distribution was accumulated near the inner disk radius
just before the burst. The mass $\delta M$ of the Gaussian distributions is  
$\sim 8\times 10^{20}$ g and $\sim 1.6\times 10^{20}$ g for the non-irradiated and 
irradiated models respectively. The amount of energy needed to push these masses from 
the inner disk radius $\Rin$ to $R_0$ obtained from the best model fits are 
$2.6 \times 10^{37}$ ergs and $5.2 \times 10^{36}$ ergs. For an isotropic burst,
the fraction of the burst energy absorbed by the disk is 
$2 \pi R (2h)/(4 \pi R^{2})=h/ R$ which is nearly constant along the disk and 
about a few $\times 10^{-3}$ for the
accretion regime of the AXP 1E 2259+58.  
Then, the required total burst energies are about $10^{40}$ ergs 
for the non-irradiated  disk model and  
$2 \times 10^{39}$ ergs for the irradiated  disk model. If the burst took place  
before the RXTE observations presented here, a burst duration more than 
about 20 seconds for the non-irradiated disk is enough for the burst luminosity 
to remain under the detection limit of the Ulysses gamma-ray detectors 
which were active before the RXTE observations (Woods et al. 2004). For the burst 
energy expected from the irradiated disk model, the burst duration must be longer 
than only $\sim 5$ seconds for the burst not to be detectable by the Ulysses 
gamma-ray detectors.

\section{Conclusion}

We have shown by means of numerical fits to data that  
the long term ($\sim 1.5$ years) contemporaneous X-ray and IR  
energy flux data of the AXP 1E 2259+58 can be 
accounted for by the evolution of the disk 
after inner disk matter is initially pushed back by a burst.
The estimated burst energy from the initial conditions of 
both irradiated and non-irradiated disk models are small enough 
to remain below the 
detection limits of the Ulysses gamma-ray detectors operating before the 
enhanced X-ray observations.

Both irradiated and non-irradiated disk models 
can reproduce the X-ray enhancement data. The irradiated disk model 
is also in agreement with the overall IR light curve accompanying the 
X-ray enhancement. The non-irradiated disk model gives a slower decay 
than that indicated by the first two IR data points corresponding to 
about three days and ten days after the onset of the X-ray enhancement. 
Nevertheless, it is consistent with the subsequent 1.5 year long term IR data.  
Since the early decay phase could depend on the details of the initial
disk conditions we can not exclude the non-irradiated disk model. 

{\bf It is not possible to compare the active and passive disk models on the basis of the  IR (K band) data (for a given X-ray light curve). Since the emission in this band comes from  the irradiation dominated outer part of the disk, the same IR radiation would be expected to be observed in either model. However, expectations of the two models are different for the shorter wavelength optical emission (R, V, B bands) and their relation to the X-ray flux. For an active disk with a mass inflow rate $\sim 10^{14} - 10^{ 15}$  g s$^ {-1}$,  significant  fraction of the emission in these optical bands comes from the intrinsic dissipation near the innermost disk radius.  Dissipation gives rise to certain amount of mass inflow, and thereby to accretion since this mass inflow is near the co-rotation radius for relevant mass inflow rates. Our model thus expects a correlation also between the dissipation dominated optical (R,V,B  bands ) luminosity and the X-ray luminosity. (Note that an inner disk emitting in the optical indicates that the dipole magnetic field is $\sim 10^{12}$ G for relevant disk mass inflow rates). Furthermore, any characteristic variation in the dissipation dominated part of the optical luminosity is expected to be followed in X-rays with a delay characterized by the viscous time scale of the inner disk (hours – days in persistent states, much shorter in the early enhancement phase due to large pressure gradients). For AXPs, in the persistent phase, our model can estimate the dissipation dominated optical flux using X-ray , IR and inter-stellar extinction  information,  which  requires a detailed source by source examination of the AXPs (work in progess).  For AXP 1E 2259+58, we estimate that the luminosities in K$_s$ and R bands are nearly the same, and  their ratios to the X-ray luminosity is about 
$10^{-4}$ in its persistent state.  At present, we cannot use this point to distinguish between our model and passive disk models like that of Rea et al.(2004), because only upper limits exist for the R and V band luminosity of AXP 2259+58. However, this remains as a distinguishing prediction of our model.}

\acknowledgments

We acknowledge support from the Astrophysics and Space
Forum at Sabanc{\i} University.
MAA acknowledges partial support from the Turkish Academy of Sciences.   
EG acknowledges support from the Turkish Academy of Sciences through a grant 
EG/T{\"U}BA-GEB{\.I}P/2004-11.


\clearpage
\begin{figure}
\vspace{-2cm}
\plotone{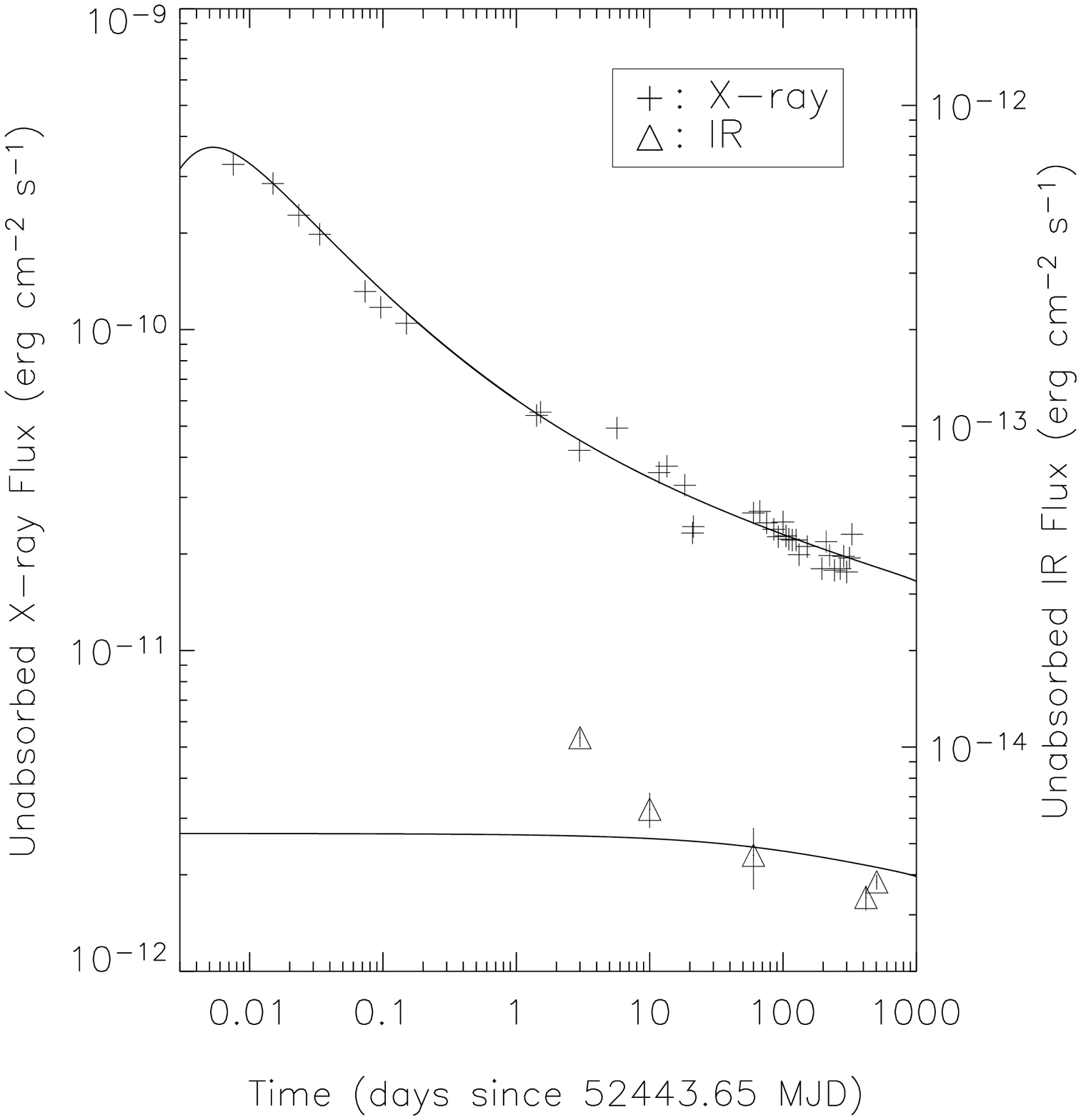} 
\vspace{0cm}
\caption{X-ray (plus signs) and IR (triangles) energy flux data of 
the enhancement phase of 1E 2259+58. The solid lines  are 
the non-irradiated disk model curves.}

\end{figure}

\clearpage
\begin{figure}
\vspace{-2cm}
\plotone{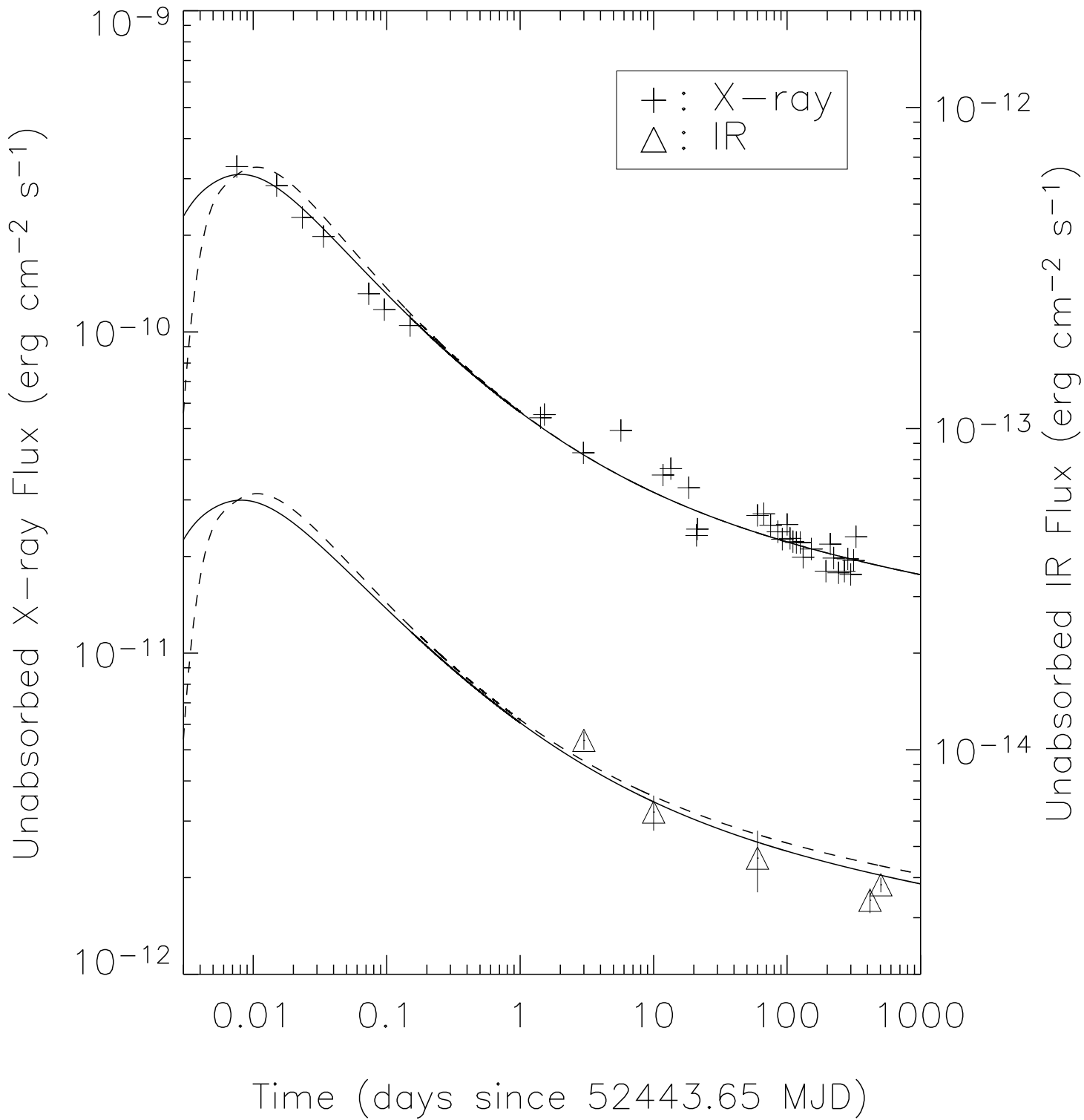} 
\vspace{0cm}
\caption{X-ray (plus signs) and IR (triangles) energy flux data of 
the enhancement phase of 1E 2259+58, and  
irradiated disk model curves. Dashed and solid curves are for two 
different initial Gaussian surface density distributions (see Table 1 for the
model parameters).}

\end{figure}

\clearpage
\begin{figure}
\vspace{-2cm}
\plotone{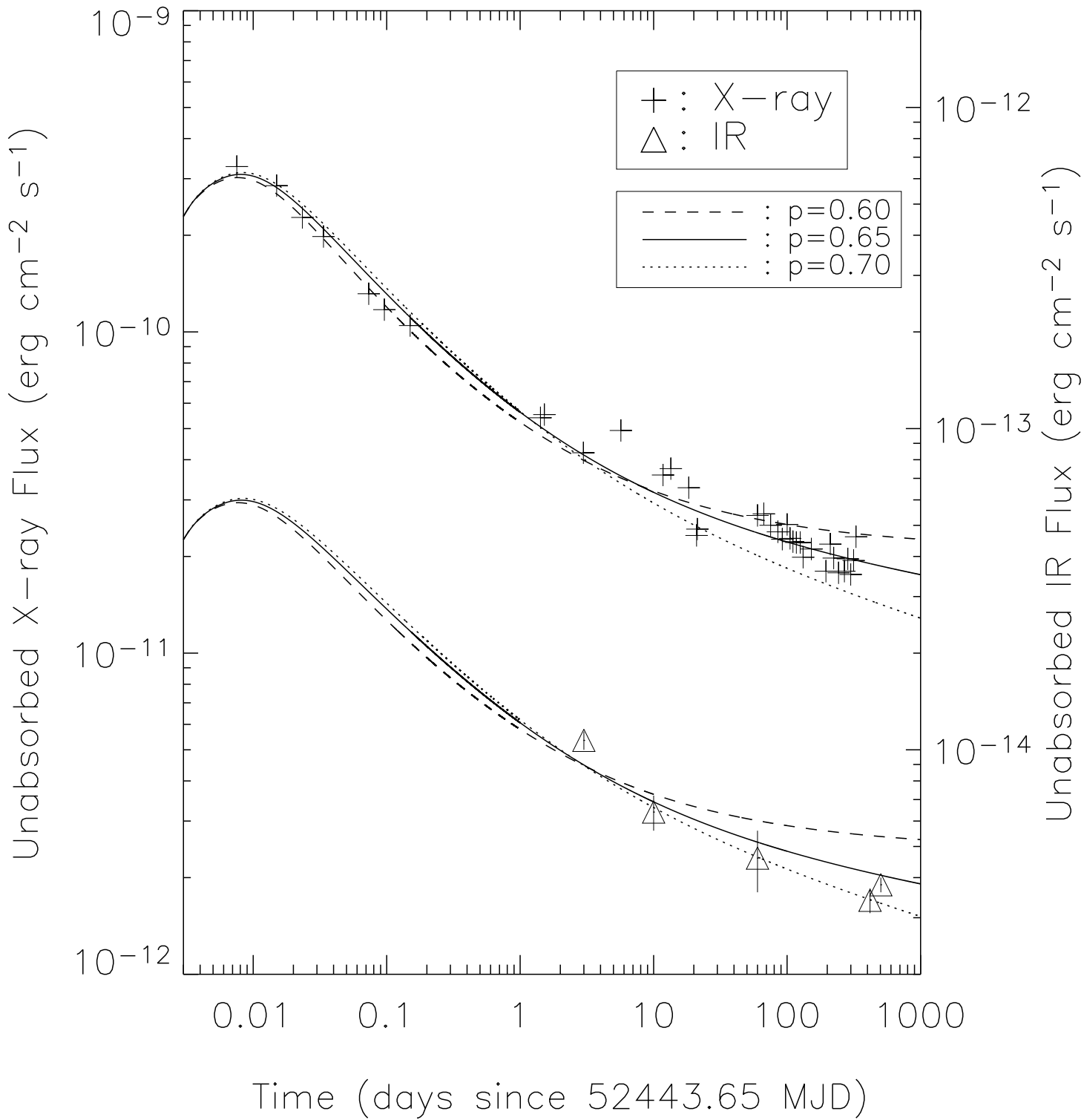} 
\vspace{0cm}
\caption{X-ray (plus signs) and IR (triangles) energy flux data of 
the enhancement phase of 1E 2259+58. The irradiated model curves correspond to 
different power law indices $p$ of the initial extended outer disk surface 
density distribution. The other disk parameters are the same for all three
curves.}

\end{figure}

\clearpage

\begin{table}
\begin{center}
\caption{Model parameters for the flux evolution presented in Figures 1 $\&$ 2. 
The three model curves in Fig. 3 have the same parameters as given in the 
first column of the irradiated disk model, but for different $p$ values (0.60, 0.65, 0.70). 
See \S 3 for the explanation of the parameters here.}

\begin{tabular}{l|c|c|c}
\tableline\tableline
&{Non-irradiated disk Model}&\multicolumn{2}{c}{Irradiated Disk Model}
\\
\hline
$\Sigma_{max}$ (g cm$^{-2}$)&$7.5\times 10^2$&$1.5\times 10^2$ 
&$3.5 \times 10^2$ 
  \\
\hline
Gaussian width (cm)&$2.0\times 10^7$ &$2.0\times 10^7$ 
&$1.1\times 10^7$ 
\\
\hline
$\Sigma_{0}/\Sigma_{\m{max}}$&0.6&0.6&0.3
\\
\hline
$R_0$ (cm)&$1.0\times 10^9$& $1.0\times 10^9$ & $1.0\times 10^9$ 
\\
\hline
$R_{\m{in}}$ (cm)&$8.3\times 10^8$ &$8.5\times 10^8$&$8.5\times 10^8$
\\
\hline
$p$ &$0.7$ &$0.65$&$0.65$  
\\
\hline
$f$ &0.03&0.27&0.27\\
\hline
$\cos i$ &0.9&0.76&0.76\\
\hline
\tableline

\end{tabular}
\end{center}
\end{table}


\begin{thebibliography}{}
\bibitem[]{571} Alpar, M.A. 2001, ApJ, 554, 1245
\bibitem[]{572} Chatterjee, P., Hernquist, L., $\&$ Narayan, R. 2000, 
ApJ, 534, 373
\bibitem[]{572} Colpi, M., Geppert, U., $\&$ Page, D. 2000, ApJ, 529, 29
\bibitem[]{572} de Jong, J. A., van Paradijs, J., $\&$ Augusteijn, T. 1996,
   A$\&$A, 314, 484
\bibitem[]{} Dhillon et al., 2005, MNRAS (in press)   
\bibitem[]{614} Dubus, G., Lasota, J.-P., Hameury, J.-M., $\&$ Charles, P, 
1999, MNRAS,
   303, 139
\bibitem[]{617} Dubus, G., Hameury, J.-M., $\&$, Lasota, J.-P. 2001, A$\&$A, 
373, 251
\bibitem[]{619} Duncan, R. C., $\&$ Thompson, C. 1992, ApJ, 392, L9
\bibitem[]{620} Ek\c{s}i, Y.K., $\&$ Alpar, M.A. 2003, ApJ, 559, 450
\bibitem[]{574} Ertan, {\"U}., $\&$  Alpar, M.A. 2003, ApJ, 593:L93
\bibitem[]{574} Ertan, {\"U}., $\&$ Cheng, K.S. 2004a, ApJ, 605, 840   
\bibitem[]{574} Ertan, {\"U}., $\&$ Cheng, K.S. 2004b, New Astronomy, 9, 503
\bibitem[]{575} Feroci, M., Hurley, K., Duncan, R., $\&$ Thompson, C. 2001,
   ApJ, 549, 1021
\bibitem[]{577} Frank, J., King, A.R., $\&$ Raine, D., 1992, Accretion 
Power in Astrophysics (Cambridge: Cambridge University Press)
\bibitem[]{630}Hulleman, F., van Kerkwijik, M. H., $\&$ Kulkarni, S.R. 2000,  
   Nature, 408, 689
\bibitem[]{632} Hulleman, F., Tennant,A.F.,van Kerkwijik,M.H.,
   Kulkarni, S.R., Kouveliotou, C., $\&$ Patel, S. K. 2001, Apj, 563, L49
\bibitem[]{634} Hulleman, F., van Kerkwijik,M.H., Kulkarni, S.R., 2004, A$\&$A, 416,   
1037
\bibitem[]{636}Hurley, K., Cline, T., Mazets, E., Barthelmy, S., 
Butterworth, P.,
   Marshall, F., Palmer, D., Aptekar, R., Golenetskii, S., Ill'lnskii, V.,
   Frederiks, D., McTiernan, J., Gold, R., $\&$ Trombka, T. 1999, 
Nature, 397, 41
\bibitem[]{641} Hurley, K. 2000 in AIP Conf. Proc. 526, Gamma-Ray Bursts:
   Fifth Huntsville Symp., ed. R. M. Kippen, R. S. Mallozzi, $\&$
   G. J. Fishman (New York: AIP), 763
\bibitem[]{644} Israel, G. L.,Covino, S., Perna, R., Mignani, R., Stella, 
L., Campana, S.,
   Marconi, G., Bono, G., Mereghetti, S., Motch, C., Negueruela, I.,
   Oosterbroek, T., $\&$ Angelini, L. 2003, ApJ, 589, L93
\bibitem[]{648} Israel, G. L.,Covino,S., Stella, L., Campana, S., Marconi, G.,
   Mereghetti, S., Mignani, R., Negueruela,I., Oosterbroek, T., Parmar, 
A. N.,
   Burderi, L., $\&$ Angelini, L. 2002, ApJ, 580, L143
\bibitem[]{652} Israel, G. L., Rea, N., Mangano, V., Testa, V., Perna, R.,
   Hummel, W., Mignani, R., Ageorges, N., Lo Curto, G., Marco, O., 
Angelini, L.,
   Campana, S., Covino, S., Marconi, G., Mereghetti, S., $\&$ Stella, L.
   2004, ApJ, 603, L97
\bibitem[]{657} Kaspi, V.M. et al. 2003, ApJ, 588, L93
\bibitem[]{658} Kern B. $\&$ Martin, C. 2002, Nature, 417, 527
\bibitem[]{581} Lyubarsky, Y., Eichler, D., $\&$ Thompson, C. 2002, ApJ, 580, L69
\bibitem[]{661} Marsden, D., Lingelfelter, R. E., Rothschild, R. E.,
    $\&$ Higdon, J. C. 2001, ApJ, 550, 397
\bibitem[]{582} Mazets, E.P., Cline, T., Aptekar, R.L., Butterworth, P.,
   Frederiks, D.D., Golenetskii, S.V., Il'inskii, V.N., $\&$ Pal'shin,
   V.D.1999, Astron.Lett., 25, 635
\bibitem[]{585} Mazets, E.P., Golenetskii, S.V., Il'inskii, V.N.,
   Aptekar, R.L., $\&$ Guryan, Y.A. 1979 Nature, 282, 587
\bibitem[]{} Menou, K., Perna, R., $\&$ Hernquist, L.  2002, ApJ, 564:L84      
\bibitem[]{669} Mereghetti, S., Chiarlone, L., Israel, G. L., $\&$ Stella, 
L. 2002,
    in Proc. 270th WE-Heraus Seminar on Neutron Stars, Pulsars and 
Supernova Remnants,
    ed. W. Becker, H. Lecsch, $\&$ J. Trumper (MPE Rep. 278; Garching: 
MPE), 29
\bibitem[]{}Palmer, D.M. et al, 2005, Nature, 434, 1107
\bibitem[]{} Rea, N. et al. 2004, A$\&$A, 425, L5    
\bibitem[]{590} Shakura, N.I., $\&$ Sunyaev, R.A., 1973, A$\&$A, 24, 337
\bibitem[]{590} Tam, C. R., Kaspi, V.M., van Kerkwijk, M.H., $\&$ 
    Durant, M. 2004, ApJ, 617, L53
\bibitem[]{591} Thompson, C., $\&$ Duncan, R. C. 1995 MNRAS, 275, 255
\bibitem[]{591} Thompson, C., $\&$ Duncan, R. C. 1996 ApJ, 473, 322
\bibitem[]{680} Thompson, C., and Duncan, C. R., Woods, P.M., Kouveliotou,
   C., Finger, M.H., $\&$ van Paradijs, J. 2000, ApJ, 543, 340  
\bibitem[]{682} Tuchman, Y., Mineshige, S. $\&$ Wheeler, J.C., 1990, ApJ, 
359, 164
\bibitem[]{591} Wang, Z., $\&$ Chakrabarty, D. 2002, ApJ, 579, 33
\bibitem[]{595} Woods, P. M., Kouveliotou, C., G\"{o}\u{g}\"{u}\c{s},
   E., Finger, M.H., Swank, J., Smith, D.A., Hurly, K., $\&$ Thompson,
   C.2001, ApJ, 552, 748
\bibitem[]{595} Woods, P. M. et al. 2004, ApJ, 605, 378 
\end{thebibliography}
\end{document}